\documentclass[aps,prb,twocolumn,superscriptaddress,showpacs]{revtex4-1}%%% <===== double-column style

\usepackage{color}

\usepackage{graphicx}

\begin{document}

\title{Data-driven determination of the spin Hamiltonian parameters and their uncertainties: \\
The case of the zigzag-chain compound KCu$_4$P$_3$O$_{12}$}

\author{Ryo Tamura}
\email[]{tamura.ryo@nims.go.jp}
\affiliation{%
International Center for Materials Nanoarchitectonics (WPI-MANA), 
National Institute for Materials Science, Ibaraki 305-0044, Japan
}%
\affiliation{%
Research and Services Division of Materials Data and Integrated System, 
National Institute for Materials Science, Ibaraki 305-0047, Japan
}%
\affiliation{%
Graduate school of Frontier Sciences, The University of Tokyo, Chiba 277-8568, Japan
}%

\author{Koji Hukushima}
\email[]{hukusima@phys.c.u-tokyo.ac.jp }
\affiliation{
Komaba Institute for Science, The University of Tokyo, 3-8-1 Komaba, Meguro, Tokyo 153-8902, Japan
}
\affiliation{%
Department of Basic Science, Graduate School of Arts and Sciences, The University of Tokyo, Komaba, Meguro, Tokyo 153-8902, Japan
}%
\affiliation{%
Research and Services Division of Materials Data and Integrated System, 
National Institute for Materials Science, Ibaraki 305-0047, Japan
}%

\author{Akira Matsuo}
\affiliation{%
The Institute for Solid State Physics, The University of Tokyo, Kashiwa, Chiba 277-8581, Japan
}%

\author{Koichi Kindo}
\affiliation{%
The Institute for Solid State Physics, The University of Tokyo, Kashiwa, Chiba 277-8581, Japan
}%

\author{Masashi Hase}
\email[]{hase.masashi@nims.go.jp}
\affiliation{%
Research Center for Advanced Measurement and Characterization, 
National Institute for Materials Science, Ibaraki 305-0047, Japan
}%

\date{\today}

\begin{abstract}

We propose a data-driven technique to estimate the spin Hamiltonian, including uncertainty, from multiple physical quantities.
Using our technique,
an effective model of KCu$_4$P$_3$O$_{12}$ is determined from the experimentally observed magnetic susceptibility and magnetization curves with various temperatures under high magnetic fields.
An effective model, which is the quantum Heisenberg model on a zigzag chain with eight spins having $J_1= -8.54 \pm 0.51 \ {\rm meV}$, 
$J_2 = -2.67 \pm 1.13 \ {\rm meV}$,
$J_3 = -3.90 \pm 0.15 \ {\rm meV}$,
and $J_4 = 6.24 \pm 0.95 \ {\rm meV}$, 
describes these measured results well. 
These uncertainties are successfully determined by the noise estimation.
The relations among the estimated magnetic interactions or physical quantities are also discussed.
The obtained effective model is useful to predict hard-to-measure properties such as spin gap, spin configuration at the ground state, magnetic specific heat, and magnetic entropy. 

\end{abstract}

%\pacs{}

%\keywords{}

\maketitle

%%%%%%%%%%%%%%%%%%%%%%%%%%%%%%%%
%Introduction
%%%%%%%%%%%%%%%%%%%%%%%%%%%%%%%%
\section{Introduction}

%Model estimation

An effective model in materials science often explains the origin of physical properties in materials.
Many methods have been developed to construct an effective model for a target material, and they can be divided into two groups. 
One is ab initio calculations, which determine the model parameters in an assumed effective model by providing only basic information of the target material\cite{Munoz-2002,Mazin-2007,Nakamura-2008,Pruneda-2008,Hirayama-2013,Nuss-2014,Misawa-2014,Riedl-2016,Hirayama-2018}.
The other is a data-driven approach in which model parameters are determined so as to fit 
the experimentally measured data in the target material\cite{Takubo-2007,Tamura-2011,DiStasio-2013,Hase-2015,Hase-2016,Hase-2017}.

%Materials informatics and effective model estimation

In the latter case,
many trials and errors are conducted to find the appropriate model parameters to describe the experimental results. 
Data-driven analyses based on machine learning have been extensively exploited to avoid this cumbersome task. 
Recently, data-driven techniques are becoming indispensable in materials science, 
because they should accelerate the discovery of novel materials\cite{Rajan-2005, Oganov-2006, Pilania-2013,Seko-2015,Ueno-2016,Ju-2017,Ramprasad-2017,Ikebata-2017,Sumita-2018,Yamashita-2018,Bombarelli-2018,Hou-2019,Terayama-2019} and deepen our understanding of materials\cite{Behler-2007,Ghiringhelli-2015,Pham-2016,Kobayashi-2017,Suzuki-2017,Dam-2018,Shiba-2018,Tamura-2019}.
From the viewpoint of effective model estimations, 
data-driven techniques are also efficient to accelerate automatic searches for appropriate model parameters\cite{Tamura-2018} and 
to extract relevant model parameters\cite{Takenaka-2014,Tamura-2017,Fujita-2018}. 

\begin{figure}[b]
\begin{center}
\includegraphics[scale=1.2]{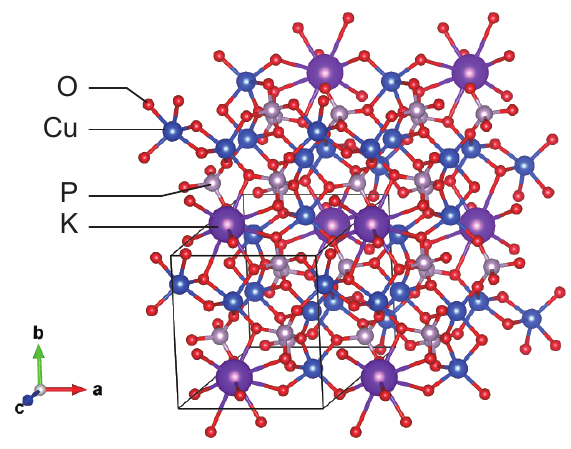} 
\end{center}
\caption{\label{fig:crystal}
(Color online)
Crystal structure of KCu$_4$P$_3$O$_{12}$ drawn by VESTA\cite{Momma-2011}.
Black box indicates a unit cell.
}
\end{figure}
%

%KCu4P3O12

The paper estimates a spin Hamiltonian as an effective model of KCu$_4$P$_3$O$_{12}$ by a data-driven approach.
Figure~\ref{fig:crystal} shows the crystal structure of KCu$_4$P$_3$O$_{12}$ ($a=$7.433 \AA, $b=$ 7.839 \AA, $c=$ 9.464 \AA, 
$\alpha=$ 108.28$^\circ$, $\beta=$ 112.68$^\circ$, $\gamma=$ 92.73$^\circ$, and space group: P$\bar{1}$)\cite{Effenberger-1987}.
Cu(II) ions have $S=1/2$ isotropic Heisenberg spin, 
but their magnetic properties have not yet been reported.
As will be explained later,
the lattice structure of the Cu ions can be regarded as a zigzag chain consisting of eight Cu ions.
Thus, the quantum Heisenberg model on a zigzag chain is the target of the spin Hamiltonian of KCu$_4$P$_3$O$_{12}$ to be estimated.
We determine the superexchange interactions between Cu ions with uncertainty in this target model by a data-driven approach in which the experimentally measured susceptibility and magnetization curves are inputted. 
Once an estimated spin Hamiltonian is established, 
theoretical analysis of the Hamiltonian  predicts various magnetic properties, which cannot or have not been measured. 
These properties include the magnetic specific heat, magnetic entropy, spin configuration, and spin gap.
These predictions are helpful to propose a further experimental plan and design.

%Organization of this paper

The rest of the paper is organized as follows.
Section II shows the experimental results of the magnetic susceptibility and magnetization curves as functions of temperatures in KCu$_4$P$_3$O$_{12}$ along with the experimental methods.
Section III explains our data-driven approach to estimate a spin Hamiltonian.
The posterior distribution of model parameters given in the data is constructed by a statistical noise model with respect to the experimental observations and the prior distribution of the model parameters.
Plausible model parameters are determined by the maximizer of the posterior distribution.
Furthermore, 
a systematic method,
which allows the statistical uncertainty of the model parameters to be evaluated,
is presented to estimate the amplitude of noise in the noise model. 
Under our formulation, multiple types of physical quantities can be used to estimate the effective model.
Section IV explains the results of the spin Hamiltonian estimation with uncertainty by our data-driven approach.
First, we assume the shape of the target spin Hamiltonian for KCu$_4$P$_3$O$_{12}$ from the viewpoint of the crystal structure.
Next, since the L$_2$ regularization is adopted as a prior distribution to suppress an increase in the absolute values of the magnetic interactions,
determination of the hyperparameter in the L$_2$ regularization is discussed.
Subsequently, 
by considering the observation noise,
four types of magnetic interactions are estimated with the error bars in a target spin Hamiltonian, 
and their relationships are discussed from the distributions of sampling data by the Monte Carlo method.
Finally,
various magnetic properties of KCu$_4$P$_3$O$_{12}$ are predicted.
Section V presents the discussion and summary.

%%%%%%%%%%%%%%%%%%%%%%%%%%%%%%%%
%Magnetic properties
%%%%%%%%%%%%%%%%%%%%%%%%%%%%%%%%
\section{Experimentally measured magnetic properties}

\begin{figure}[]
\begin{center}
\includegraphics[scale=0.45]{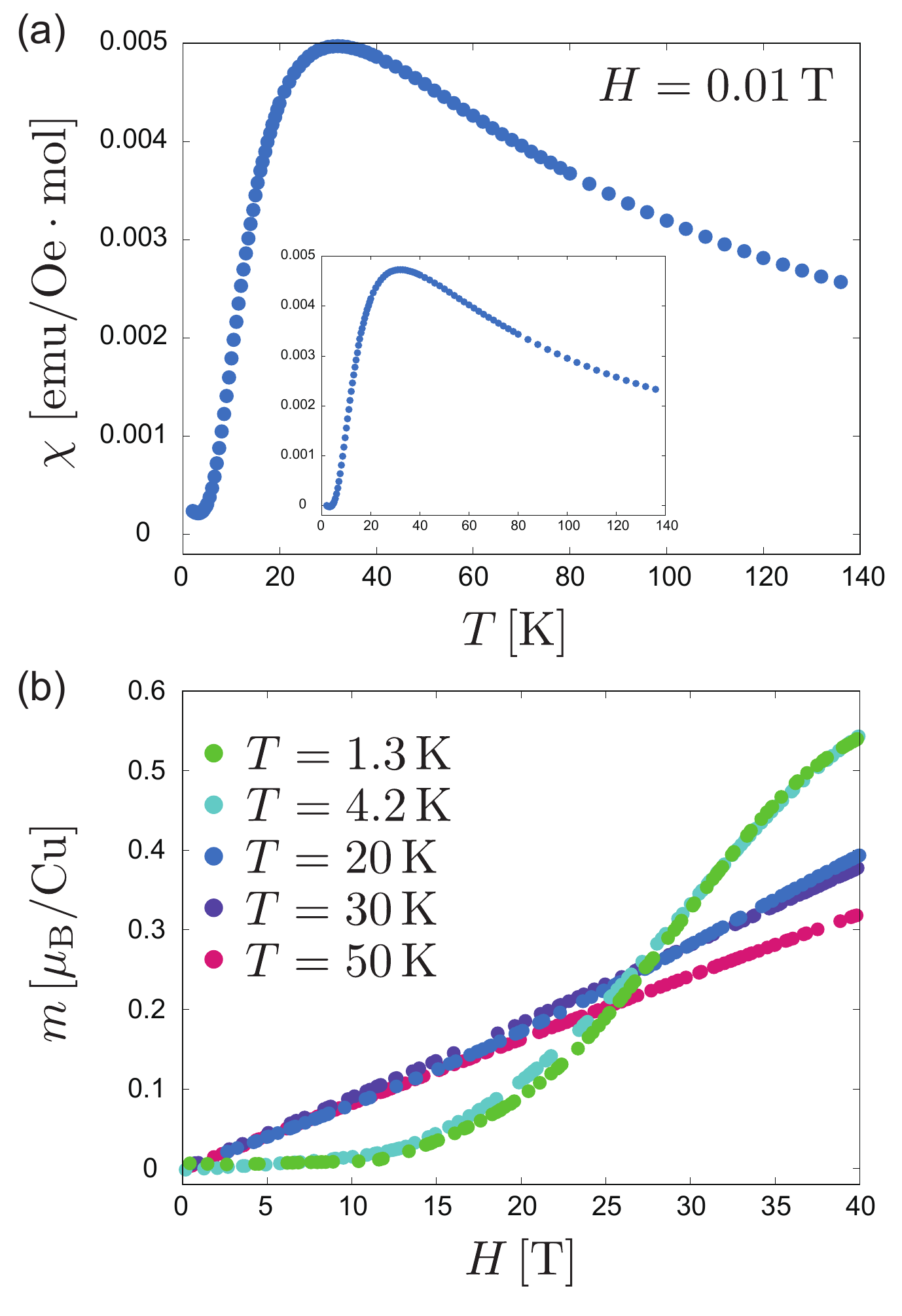} 
\end{center}
\caption{\label{fig:experiments}
(Color online)
(a) Magnetic susceptibility with 0.01 T for KCu$_4$P$_3$O$_{12}$.
Inset shows the results where the constant term in the susceptibility is removed.
(b) Magnetization curves at various temperatures for KCu$_4$P$_3$O$_{12}$.
}
\end{figure}

Figure~\ref{fig:experiments} (a) shows the temperature dependence of the magnetic susceptibility with a magnetic field of 0.01 T,
which was measured by a superconducting quantum interference device magnetometer, magnetic property measurement system (Quantum Design).
Crystalline KCu$_4$P$_3$O$_{12}$ powder was synthesized by a solid-state reaction.
Even at sufficiently low temperatures,
a finite constant value of susceptibility remains.
The inset of Fig.~\ref{fig:experiments} (a) shows the susceptibility upon removing this constant term. 
This data is used in the data-driven approach.
Figure~\ref{fig:experiments} (b) shows the magnetization curves at temperatures of 1.3 K, 4.2 K, 20 K, 30 K, and 50 K.
These magnetization curves were measured using an induction method with a multilayer pulsed field magnet installed at the Institute for Solid State Physics, the University of Tokyo.
Although a high magnetic field is imposed ($\le 40$ T),
the magnetization is not saturated.
From the ESR measurements,
the $g$-factor of Cu ions is determined to be 2.08.

%

%%%%%%%%%%%%%%%%%%%%%%%%%%%%%%%%
%Method
%%%%%%%%%%%%%%%%%%%%%%%%%%%%%%%%
\section{Data-driven approach to estimate an effective model with uncertainty}

\subsection{Posterior distribution for effective model estimation}

Our developed effective model estimation method in Ref.~\onlinecite{Tamura-2017} is based on the Bayesian statistics.
We consider that the target Hamiltonian to be estimated has $K$-types of model parameters: $\mathbf{x} = (x_1, ..., x_K)$.
Let $\mathbf{y}^{\rm ex} = \{ y^{\rm ex}(g_l) \}_{l=1, ..., L}$ be the set of experimentally measured physical quantities. 
This set depends on the external parameter $g_l$, where the number of data is $L$.
Using Bayes' theorem,
the posterior distribution $P(\mathbf{x}| \mathbf{y}^{\rm ex})$,
which is the conditional probability of the model parameters given experimental data,
is expressed as
\begin{eqnarray}
P(\mathbf{x}| \mathbf{y}^{\rm ex} ) \propto P( \mathbf{y}^{\rm ex} |\mathbf{x}) P(\mathbf{x}),
\label{eq:Bayes_theorem}
\end{eqnarray}
where $P(\mathbf{x})$ is the prior distribution of the model parameters and $P(\mathbf{y}^{\rm ex} |\mathbf{x})$ is the likelihood function of $\mathbf{y}^{\rm ex}$ given $\mathbf{x}$. 
Assuming that the observation noise for the measurements follows a Gaussian distribution with a mean of zero and a standard deviation of $\sigma$, 
the likelihood function is given by 
\begin{eqnarray}
&&P(\mathbf{y}^{\rm ex} |\mathbf{x}) \nonumber \\
&&= \left( \frac{1}{2 \pi \sigma^2} \right)^\frac{L}{2} \exp \left[ - \frac{1}{2 \sigma^2} \sum_{l=1}^L \left( y^{\rm ex}(g_l) - y^{\rm cal} (g_l, \mathbf{x}) \right)^2 \right]. \nonumber \\
\label{eq:cond_phys_qu}
\end{eqnarray}
Here, $\{ y^{\rm cal} (g_l, \mathbf{x}) \}_{l=1, \cdots, L}$, which is expressed as $\mathbf{y}^{\rm cal} (\mathbf{x})$, 
are the physical quantities calculated from the Hamiltonian at the external parameter $g_l$ theoretically. 

To express the posterior distribution briefly,
we introduce the ``energy function'' as a function of $\mathbf{x}$ and the noise $\sigma$ is given by
\begin{equation}
E (\mathbf{x},\sigma) = \Delta (\mathbf{x},\sigma) - \log P (\mathbf{x}), \label{eq:energy_function}     
\end{equation}
where an error function $\Delta (\mathbf{x},\sigma)$  is 
\begin{equation}
\Delta (\mathbf{x},\sigma) = \frac{1}{2 \sigma^2} \sum_{l=1}^L \left( y^{\rm ex}(g_l) - y^{\rm cal} (g_l, \mathbf{x}) \right)^2.     
\end{equation}
Then, 
the posterior distribution is expressed as
\begin{eqnarray}
P(\mathbf{x}|\mathbf{y}^{\rm ex}) \propto \left( \frac{1}{2 \pi \sigma^2} \right)^\frac{L}{2} \exp \left[ -E(\mathbf{x},\sigma)\right].  \label{eq:cond_model}
\end{eqnarray}
From the viewpoint of the maximum a posterior (MAP) estimation,
plausible model parameters to explain $\mathbf{y}^{\rm ex}$ are regarded as the maximizer of Eq.~(\ref{eq:cond_model}). 
Correspondingly, 
the MAP estimation is reduced to a minimization problem of the energy function. 
To search for the maximizer of Eq.~(\ref{eq:cond_model}) or the minimizer of Eq.~(\ref{eq:energy_function}), 
various optimization techniques such as the steepest-descent method, Markov chain Monte Carlo (MCMC) method\cite{Tamura-2017}, and Bayesian optimization\cite{Tamura-2018} can be used.

Selecting the prior distribution of the model parameters $P(\mathbf{x})$ is important to obtain a physically appropriate Hamiltonian\cite{Tamura-2017}. 
The prior distribution for the effective model estimation can be regarded as the regularization terms in the minimization problem\cite{Bishop-2006}. 
The most common are L$_1$ (LASSO) and L$_2$ (ridge) regularization with corresponding prior distributions $P(\mathbf{x}) = \exp (-\lambda |\mathbf{x}|)$ and  $P(\mathbf{x}) = \exp (-\lambda \|\mathbf{x}\|^2)$, respectively, 
where hyperparameter $\lambda$ determines the strength of regularization.
If the L$_1$ regularization is applied, model parameters with large contributions can be selected based on the feature selection.
On the other hand, 
the L$_2$ regularization suppresses the increase in the absolute values of the model parameters.
Depending on the situation and purpose, it is necessary to select the proper prior distribution.

\subsection{Observation noise estimation}

To obtain the uncertainty of the model parameters by a MAP estimation,
the width of the posterior distribution around the maximizer must be estimated. 
The noise amplitude $\sigma$ is crucial to estimate the uncertainty\cite{Anada-2017,Tokuda-2017,Obinata-2019,Katakami-2019}. 
In our framework, 
the noise amplitude $\sigma$ is considered to be
a hyperparameter and a plausible value is determined by minimizing the Bayes free-energy $F(\sigma)$ defined as
\begin{equation}
F(\sigma) := - \log Z (\sigma),     
\end{equation}
where $Z (\sigma)$ is the normalization of the posterior distribution given by 
\begin{equation}
Z (\sigma)
=\left( \frac{1}{2 \pi \sigma^2} \right)^\frac{L}{2} \int_{\Omega_x} d \mathbf{x} \exp \left[ -E(\mathbf{x},\sigma)\right],  
\end{equation}
where $\Omega_x$ is the support of the posterior distribution determined by the prior distribution. 
To evaluate $F(\sigma)$, 
it is convenient to extend the posterior distribution $P(\mathbf{x}|\mathbf{y}^{\rm ex})$  to $P_\beta(\mathbf{x}|\mathbf{y}^{\rm ex})$ with a ``finite temperature'', 
which is defined as
\begin{eqnarray}
P(\mathbf{x}|\mathbf{y}^{\rm ex}) := \frac{1}{Z_\beta (\sigma)}\left( \frac{1}{2 \pi \sigma^2} \right)^\frac{L}{2}
\exp \left[ - \beta E(\mathbf{x},\sigma)\right],
\label{eq:prior-beta}
\end{eqnarray}
where $\beta$ is the inverse temperature and $Z_\beta(\sigma)$ is the normalization. 
By using $Z_\beta (\sigma)$,
the Bayes free-energy is calculated as
\begin{eqnarray}
F(\sigma) &=& -\int_0^1 d \beta \left( \frac{d}{d \beta} \log Z_\beta (\sigma)\right) - \log Z_0 (\sigma) \nonumber \\
&=& \int_0^1 d \beta \langle E (\mathbf{x},\sigma) \rangle_{\beta} + \frac{L}{2} \log \left( 2 \pi \sigma^2 \right) - \log \int_{\Omega_x} d \mathbf{x}, \nonumber \\
\label{eq:logZ1}
\end{eqnarray}
where the ensemble average $\langle E (\mathbf{x},\sigma) \rangle_{\beta}$ with respect to Eq.~(\ref{eq:prior-beta}) can be obtained by the MCMC method. 
Here, the third term in RHS of Eq.~(\ref{eq:logZ1}) does not depend on $\sigma$ 
and is omitted below. 
The noise amplitude of the experimental data $\sigma^*$ is evaluated as the value where $F(\sigma)$ is minimized. 
By MCMC sampling from Eq.~(\ref{eq:cond_model}) with the fixed $\sigma^*$,
the uncertainty of the model parameters can be evaluated.

\subsection{Multiple sets of physical quantities}

In our experiments of KCu$_4$P$_3$O$_{12}$,
six different sets of physical quantities, 
that is, the susceptibility and the magnetization curves under different temperatures, are obtained. 
Then different types of physical measurements are likely to be combined in our estimation problem. 

For simplicity, 
the different physical quantities are assumed to be independently obtained.
Then the likelihood function of a series of experimental results is defined as
\begin{eqnarray}
P(\mathbf{y}_1^{\rm ex}, \mathbf{y}_2^{\rm ex}, ..., \mathbf{y}_N^{\rm ex} |\mathbf{x}) \propto \prod_{n=1}^N P(\mathbf{y}_n^{\rm ex} |\mathbf{x}),
\end{eqnarray}
where the index $n$ denotes the type of physical quantities,
and $N$ is the total number of types. 
Thus, the posterior distribution is written as
\begin{eqnarray}
&&P(\mathbf{x}|\mathbf{y}_1^{\rm ex}, \mathbf{y}_2^{\rm ex}, ..., \mathbf{y}_N^{\rm ex}) \nonumber \\
&&\propto 
\prod_{n=1}^N \left( \frac{1}{2 \pi \sigma_n^2} \right)^{\frac{L_n}{2}} \exp \left[ - E (\mathbf{x},\sigma_1,...,\sigma_N) \right], 
\end{eqnarray}
where $L_n$ is the number of data points for the $n$-th type measurement and the energy function is a weighted sum given as
\begin{eqnarray}
&&E (\mathbf{x},\sigma_1,...,\sigma_N) \nonumber \\
&&=\sum_{n=1}^N \left[ \frac{1}{2 \sigma_n^2} \sum_{l=1}^{L_n} \left( y_n^{\rm ex}(g_l) - y_n^{\rm cal} (g_l, \mathbf{x}) \right)^2 \right]-\log P(\mathbf{x}). \nonumber \\
\end{eqnarray}
This means that the posterior distribution significantly depends on both  $L_n$ and $\sigma_n$. 

In this paper, for simplicity, the number of data points for all inputted physical quantities is arranged as $L_n = L$ for $\forall n$.
Furthermore,
the case where $\sigma_n$ does not depend on the type of physical quantities is considered.
That is,
the standard deviation of the observation noise is the same for all types of physical quantities:
$\sigma_n = \sigma$ for $\forall n$ .
To make this assumption more realistic,
the contributions of each type of physical quantity are arranged and the following normalization is imposed:
\begin{eqnarray}
y_n^{\rm ex}(g_l) &\to& \frac{y_n^{\rm ex}(g_l)- \min_l (y_n^{\rm ex}(g_l))}{\max_l (y_n^{\rm ex}(g_l))-\min_l (y_n^{\rm ex}(g_l))}, \\
y_n^{\rm cal}(g_l) &\to& \frac{y_n^{\rm cal}(g_l)- \min_l (y_n^{\rm ex}(g_l))}{\max_l (y_n^{\rm ex}(g_l))-\min_l (y_n^{\rm ex}(g_l))}.
\end{eqnarray}
By using the normalization,
the upper and lower bounds for each type of physical quantity are 1 and 0, respectively.

%%%%%%%%%%%%%%%%%%%%%%%%%%%%%%%%
%Results
%%%%%%%%%%%%%%%%%%%%%%%%%%%%%%%%
\section{Spin Hamiltonian estimation}

\subsection{Target Hamiltonian}

\begin{figure}[]
\begin{center}
\includegraphics[scale=0.8]{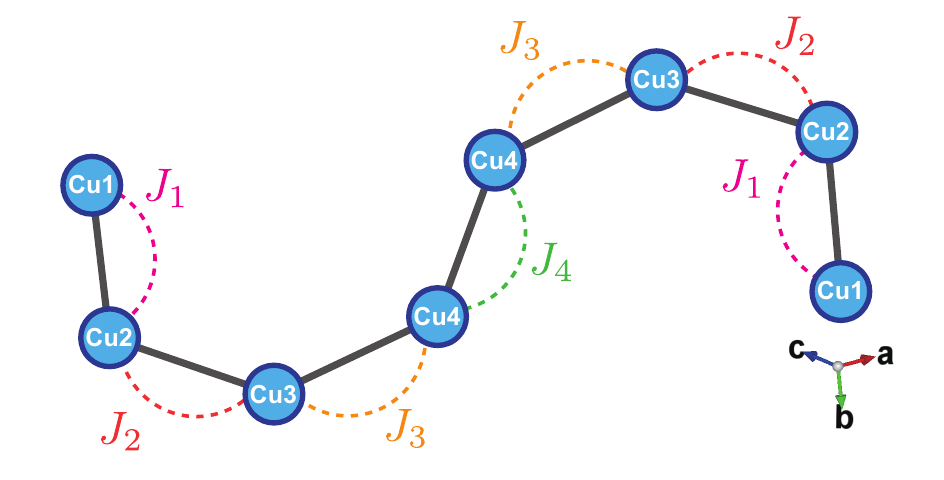} 
\end{center}
\caption{\label{fig:lattice_eff}
(Color online)
Lattice structure of the target Hamiltonian for KCu$_4$P$_3$O$_{12}$.
This zigzag chain is constructed by eight Cu ions.
Cu ions with the same index have equivalent positions when considering symmetry.
}
\end{figure}

Our model estimation method assumes the shape of the target Hamiltonian to be estimated.
KCu$_4$P$_3$O$_{12}$ has a complicated three-dimensional structures (Fig.~\ref{fig:crystal}).
On the other hand,
by only focussing on the superexchange interactions, 
that is, the Cu--O--Cu paths in the crystal structure,
the lattice structure of the Cu ions is approximated by a set of independent zigzag chains.
Each chain has eight Cu ions, and the bond lengths between the Cu ions are smaller than 3.2 \AA.
Note that this approximation should be valid except for extremely low temperatures.
For low temperatures,
weak interactions besides these superexchange interactions may affect magnetism\cite{Hase-2016,Hase-2017,Hase-2018},
and this approximation is poor.
In this paper, 
by focussing on the magnetic properties except for those at low temperatures,
the target Hamiltonian is set to a Heisenberg model on the chain with eight spins.
Figure~\ref{fig:lattice_eff} shows the lattice structure.

Although symmetry considerations imply that there are seven types of nearest neighbor interactions in each chain,
only four types of independent interactions appear
(i.e., $J_1$, $J_2$, $J_3$, and $J_4$ in Fig.~\ref{fig:lattice_eff}).
Furthermore, since the magnetic ion is Cu, 
anisotropy should not be considered.
Consequently, the target Hamiltonian with isotropic Heisenberg spins is written as
\begin{eqnarray}
\mathcal{H} (\mathbf{x}) = - \sum_{i=1}^7 J_{i,i+1} (\hat{s}^x_i \hat{s}^x_{i+1} + \hat{s}^y_i \hat{s}^y_{i+1} + \hat{s}^z_i \hat{s}^z_{i+1}),
\end{eqnarray}
where $J_1 = J_{1,2}=J_{7,8}$, $J_2 = J_{2,3}=J_{6,7}$, $J_3 = J_{3,4}=J_{5,6}$, and $J_4 = J_{4,5}$.
Here, $(\hat{s}^x_i, \hat{s}^y_i, \hat{s}^z_i)$ are the $S=1/2$ Pauli matrices on $i$th site.
The magnetic properties of this target quantum Hamiltonian can be easily calculated by the exact diagonalization method.
If an effective model with a large number of spins need to be estimated,
another simulation method such as the Monte Carlo method or mean-field calculation should be performed.

Since many magnetic interactions are already trimmed,
there is no need to use the L$_1$ regularization for the prior distribution.
On the other hand,
large magnetic interactions are not preferable for the physical sense.
To avoid an increase in the absolute values of the estimated magnetic interactions,
we adopt the L$_2$ regularization as the prior distribution.
Furthermore,
to estimate an effective model to roughly capture the magnetic properties under wide temperature and magnetic field ranges,
six types of physical quantities ($N=6$) measured in KCu$_4$P$_3$O$_{12}$ (see Fig.~\ref{fig:experiments}) are used as inputted data.
In addition, the number of data points in each inputted physical quantity is fixed as $L=100$.

\subsection{Determination of hyperparameter in the L$_2$ regularization}

To determine the hyperparameter $\lambda$ in the L$_2$ regularization,
the maximizer of the posterior distribution is analyzed.
Since $\sigma$ does not generally depend on the value of the model parameters from the viewpoint of a MAP estimation,
we search for the minimizer of the following equations:
\begin{eqnarray}
E' (\mathbf{x},\alpha) &=& \Delta' (\mathbf{x}) + \alpha \| \mathbf{x} \|^2, \\
\Delta' (\mathbf{x}) &=& \sum_{n=1}^N \sum_{l=1}^{L} \left( y_n^{\rm ex}(g_l) - y_n^{\rm cal} (g_l, \mathbf{x}) \right)^2,
\end{eqnarray}
where $\alpha := 2 \sigma^2 \lambda$, and $E' (\mathbf{x},\alpha)$ and $\Delta' (\mathbf{x})$ are the energy function and error function normalized by $1/2 \sigma^2$.
Note that determining the plausible value of $\alpha$ is equivalent to deciding the hyperparameter $\lambda$ in the L$_2$ regularization depending on $\sigma$.

\begin{figure}[]
\begin{center}
\includegraphics[scale=0.8]{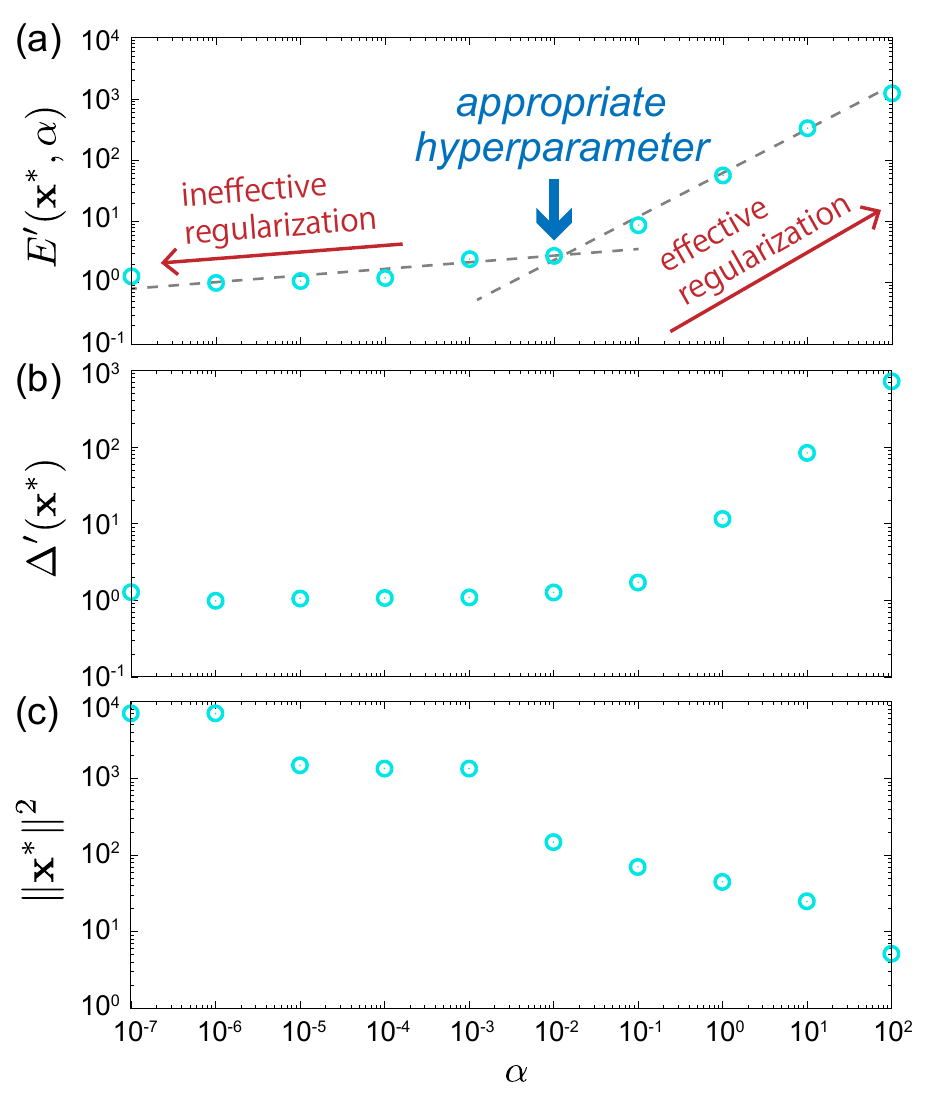} 
\end{center}
\caption{\label{fig:all_search}
(Color online)
(a) Hyperparameter $\alpha$ dependence of $E'(\mathbf{x}^*,\alpha)$ where $\mathbf{x}^*$ is the magnetic interactions so that $E'(\mathbf{x},\alpha)$ is minimized with a fixed $\alpha$.
Dotted lines are a visual guide.
(b) $\alpha$ dependence of $\Delta'(\mathbf{x}^*)$. 
(c) $\alpha$ dependence of $\|\mathbf{x}^*\|^2$.
}
\end{figure}

The minimizer of $E' (\mathbf{x},\alpha)$ is searched by the MCMC method where the probability distribution is proportional to $\exp [-E' (\mathbf{x},\alpha)]$. 
The MCMC samplings are performed by emcee package\cite{emcee,Foreman-Mackey-2012} for various $\alpha$.
In each MCMC sampling, the stretch move\cite{Goodman-2010} is used for the update scheme of states, and 8,000 states are sampled.
Among the sampled states with fixed $\alpha$,
the model parameters $\mathbf{x}^*$ such that $E'(\mathbf{x},\alpha)$ is minimized are selected.
As mentioned in Sec. III A,
this optimization of $E'(\mathbf{x},\alpha)$ can be also performed by various fast optimization techniques instead of MCMC.
Figure~\ref{fig:all_search} shows the $\alpha$ dependence of $E'(\mathbf{x}^*,\alpha)$, $\Delta' (\mathbf{x}^*)$, and $\|\mathbf{x}^*\|^2$.
Here, all experimental results, that is, the magnetic susceptibility with 0.01 T and magnetization curves at five temperatures (1.3 K, 4.2 K, 20 K, 30 K, and 50 K), for KCu$_4$P$_3$O$_{12}$ are inputted.
The result of $E'(\mathbf{x}^*,\alpha)$ shows an elbow curve, 
indicating a boundary between regions, where in each part, the regularization is effective or ineffective.
Thus, the appropriate $\alpha$ is selected as the elbow position.
That is, $\alpha^* = 10^{-2}$,
because we want to find the spin Hamiltonian not only to explain the experimental results but also to obtain small magnetic interactions as possible.
This is similar to the determination technique of the cluster number in clustering analysis\cite{Thorndike-1953,Tibshirani-2001}.

In the error function $\Delta' (\mathbf{x}^*)$,
the elbow curve is also observed,
but the elbow position deviates in $E'(\mathbf{x}^*,\alpha)$.
Preferably, 
$\Delta' (\mathbf{x}^*)$ is a small value at $\alpha^*=10^{-2}$,
and if the value of $\alpha$ decreases, 
the error is almost unchanged.
In addition, $\|\mathbf{x}^*\|^2$ monotonically decreases against $\alpha$,
and it becomes sufficiently small at $\alpha^*=10^{-2}$.
In this stage,
the estimated magnetic interactions (i.e., $\mathbf{x}^*$) at $\alpha^*=10^{-2}$ are $J_1 = -6.65$ meV, $J_2 = -5.49$ meV, $J_3 = -4.96$ meV, and $J_4 = 6.92$ meV.
The absolute values of these interactions are proper in the Cu systems\cite{Hase-2016,Hase-2017}.
These facts indicate that the L$_2$ regularization is useful to estimate a spin Hamiltonian with small magnetic interactions.

\subsection{Evaluation of the uncertainty and distribution of sampling data}

\begin{figure}[]
\begin{center}
\includegraphics[scale=0.8]{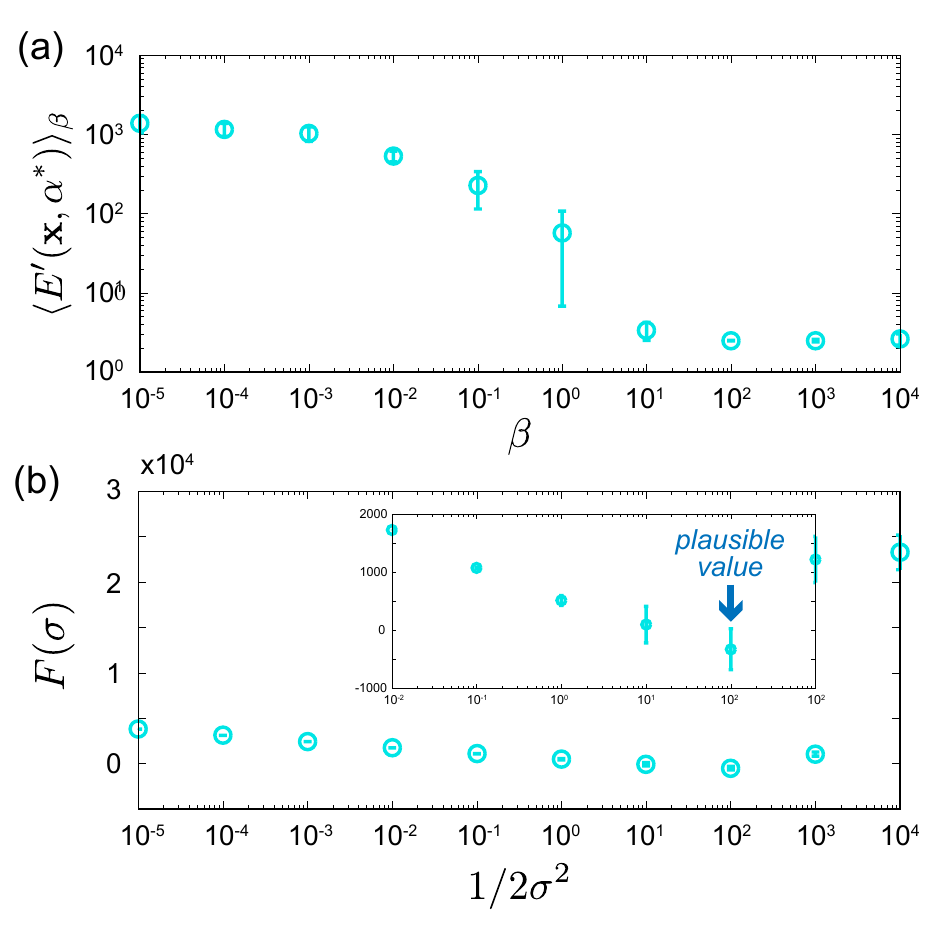} 
\end{center}
\caption{\label{fig:noise}
(Color online)
(a) Inverse temperature $\beta$ dependence of $\langle E' (\mathbf{x},\alpha^*) \rangle_{\beta}$ by MCMC.
(b) Bayes free-energy dependence on the noise amplitude $\sigma$.
Inset is the enlarged view.
}
\end{figure}

To evaluate the uncertainty of the estimated magnetic interactions,
the noise amplitude is assessed according to Sec.~III B.
Here, the Bayes free-energy in the estimation problem for KCu$_4$P$_3$O$_{12}$ is given by
\begin{eqnarray}
F(\sigma) = \int_0^\frac{1}{2\sigma^2} d \beta \langle E' (\mathbf{x},\alpha^*) \rangle_{\beta} + \frac{NL}{2} \log \left( 2 \pi \sigma^2 \right),
\end{eqnarray}
where $N=6$ and $L=100$.
Figure~\ref{fig:noise} (a) shows the inverse temperature $\beta$ dependence of $\langle E' (\mathbf{x},\alpha^*) \rangle_{\beta}$ by MCMC calculations using emcee package where the Monte Carlo step is 8,000.
The average values of eight independent runs are plotted, and the error bars denote the 95\% confident intervals.
It monotonically decreases as a function against $\beta$.
Using the results of $\langle E' (\mathbf{x},\alpha^*) \rangle_{\beta}$, 
Fig.~\ref{fig:noise} (b) shows the $\sigma$ dependence of $F(\sigma)$ which is calculated by numerical integration with the trapezoidal rule.
The minimum value of $F(\sigma)$ is obtained at $2 \sigma^{*2} = 10^{-2}$.
This value is a plausible standard deviation for the observation noise.
This means that the noise amplitude is $\sim$ 7\% when the six types of physical quantities are inputted.

\begin{figure}[]
\begin{center}
\includegraphics[scale=0.45]{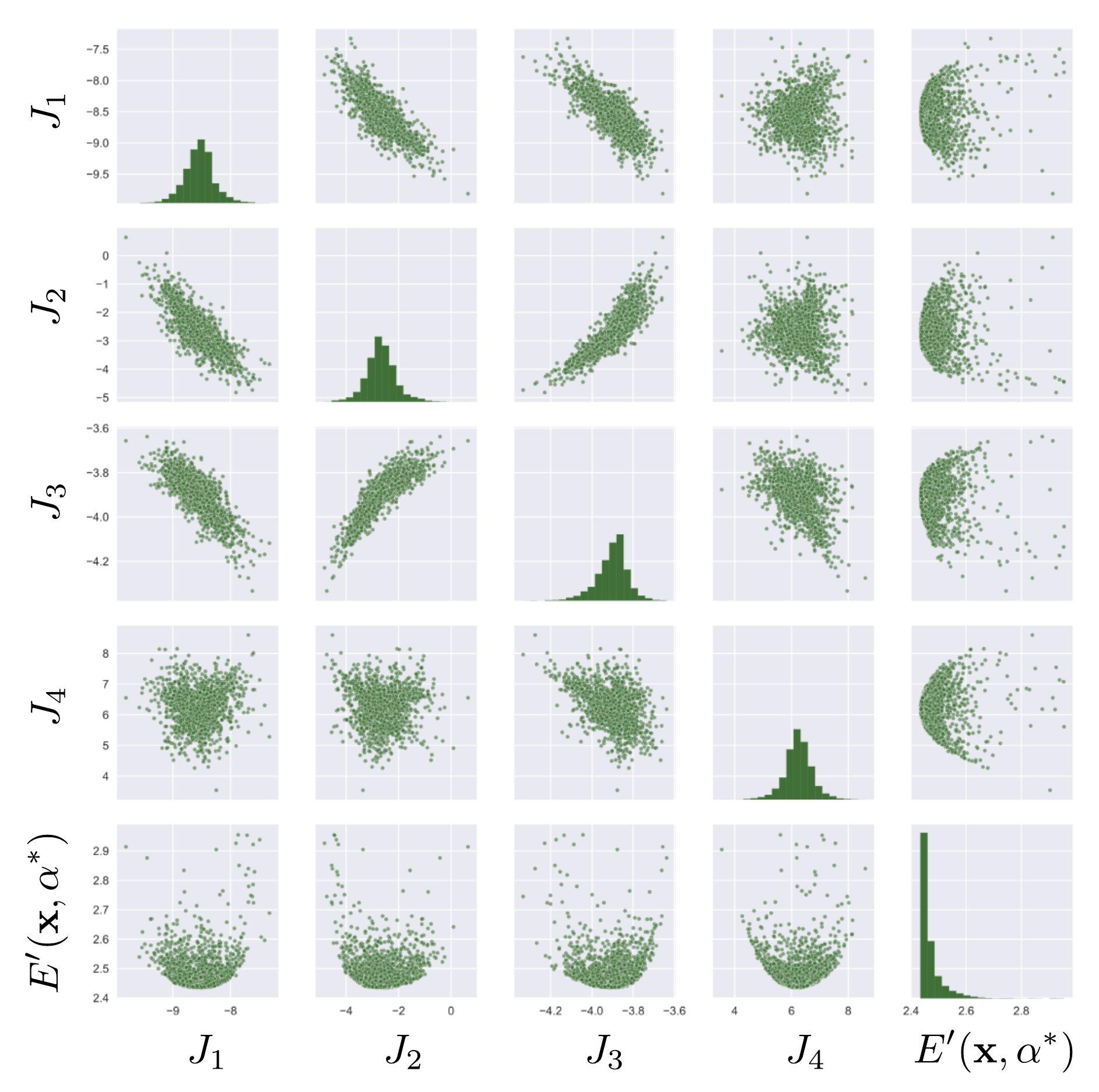} 
\end{center}
\caption{\label{fig:scatter}
(Color online)
Scatterplots of the sampling results between all combinations of $J_1$, $J_2$, $J_3$, $J_4$, and $E' (\mathbf{x},\alpha^*)$ by MCMC around the estimated magnetic interactions.
Here, the probability distribution in MCMC is proportional to $\exp (-E' (\mathbf{x},\alpha^*)/2 \sigma^{*2})$ with $2 \sigma^{*2}=10^{-2}$ and $\alpha^* = 10^{-2}$.
}
\end{figure}

Using the appropriate observation noise,
MCMC sampling is performed around the estimated magnetic interactions in Sec. IV B when the probability distribution is proportional to $\exp (-E' (\mathbf{x},\alpha^*)/2 \sigma^{*2})$ with $2 \sigma^{*2}=10^{-2}$ and $\alpha^* = 10^{-2}$.
Figure~\ref{fig:scatter} shows the scatterplots of the sampling results between all combinations of $J_1$, $J_2$, $J_3$, $J_4$, and $E (\mathbf{x})$.
Here, the last 3,000 sampling results in the MCMC with 8,000 Mote Carlo steps are plotted.
First, we determine the estimated magnetic interactions and the error bars as 95 \% confidence interval are evaluated by these distributions as
\begin{eqnarray}
J_1 &=& -8.54 \pm 0.51 \ {\rm meV}, \label{eq:MI1} \\
J_2 &=& -2.67 \pm 1.13 \ {\rm meV}, \label{eq:MI2} \\
J_3 &=& -3.90 \pm 0.15 \ {\rm meV}, \label{eq:MI3} \\
J_4 &=& 6.24 \pm 0.95 \ {\rm meV}. \label{eq:MI4}
\end{eqnarray}
Note that the inputted experimental results (Fig.~\ref{fig:experiments}) are not noisy,
and the evaluated error bars for the estimated interactions are sufficiently small.
Thus, to consider more noisy cases,
the artificial noises are added to the experimental results of KCu$_4$P$_3$O$_{12}$.
Figures S1 and S2 in supplemental material show the estimation results depending on the artificial noise.
We confirm that if the artificial noise is increased,
the value of $\sigma^*$ by our noise estimation is also increased and consequently the error bars for the estimated interactions become large.
Thus, we conclude that our estimation method can be applied to the noisy experimental results and correctly evaluate the uncertainty in the estimated effective model.

Next, from the distributions of magnetic interactions (Fig.~\ref{fig:scatter}),
the region of $J_3$ realizing good fitting is narrower compared to the other interactions.
That is,
$J_3$ is the most sensitive.
In contrast,
changing $J_2$ has a smaller impact on the fitting error,
indicating $J_2$ has the highest uncertainty.
Furthermore, we can see that $J_2$ and $J_3$ are positively correlated,
while $J_1$ negatively is correlated with $J_2$ and $J_3$.
If $J_1$ increases, $J_2$ and $J_3$ should decrease to maintain the good fitting.
On the other hand, 
$J_4$ is almost independent of the magnetic interactions.
Consequently,
$J_4$ can be freely tuned within the error bar.
We discuss these extracted correlations from physical insights.
Considering the lattice symmetry, $J_2$ and $J_3$ are placed in a similar environment, for example, 
there are two places in the lattice and the interacted spins are not edge one. 
Thus, $J_2$ and $J_3$ would have a positive correlation. 
Furthermore, $J_1$ is an antiferromagnetic interaction as well as $J_2$ and $J_3$, 
and $J_1$ should become smaller with increasing $J_2$ and $J_3$ to keep the energy scale of antiferromagnetic interactions in the Hamiltonian,
which means that $J_1$ is negatively correlated with $J_2$ and $J_3$.
We note that these correlations obtained in the estimation strongly depend on the target Hamiltonian and estimated interaction parameters.
Thus, these discussions are correlations of the estimated interaction parameters in the effective model, not correlations of the magnetic interactions in the real materials.
However, these extracted correlations can deepen an understanding of properties of the estimated effective model.

\begin{figure*}[]
\begin{center}
\includegraphics[scale=0.37]{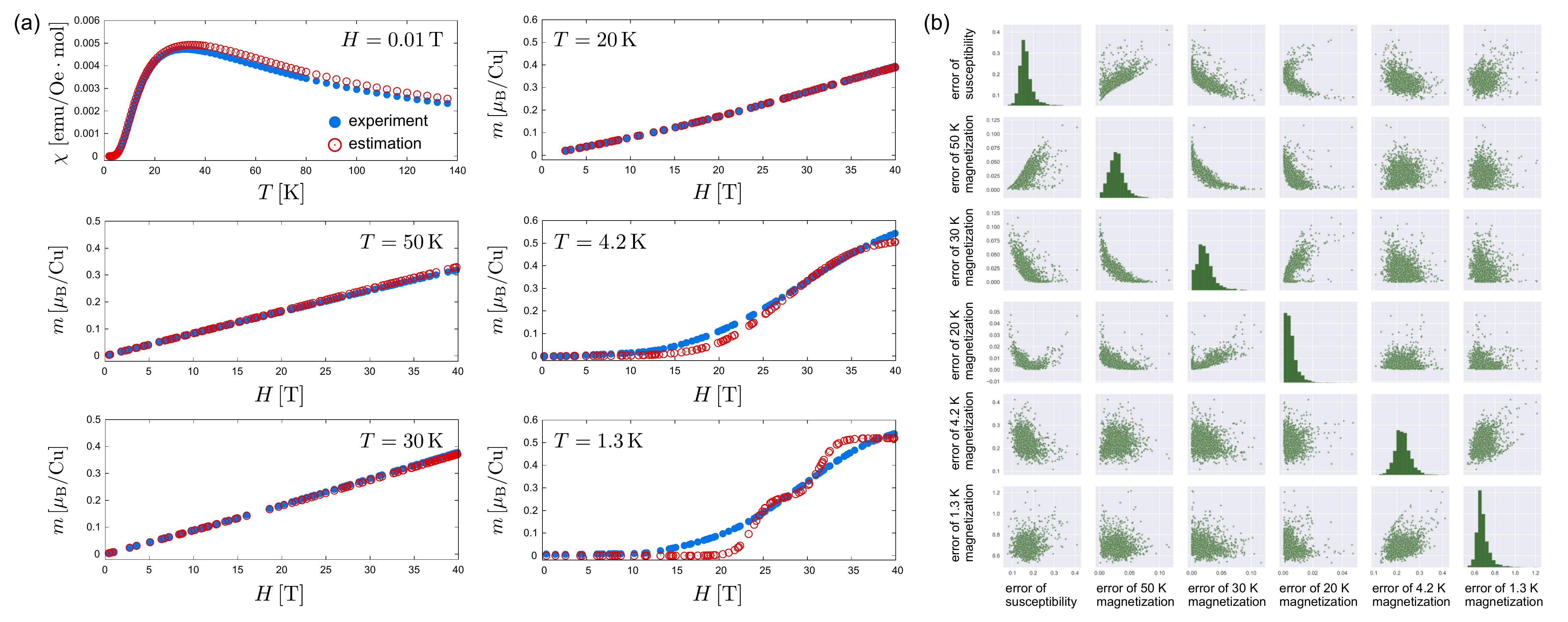} 
\end{center}
\caption{\label{fig:mag_sus_phys_scat}
(Color online)
(a) Comparison plots of the magnetic susceptibility with 0.01 T and the magnetization curves with various temperatures between the experimental and calculated results by the estimated spin Hamiltonian.
(b) Scatterplots of the errors between the experimental and calculated results,
which are independently evaluated in the susceptibility and five types of magnetization curves.
}
\end{figure*}

Figure~\ref{fig:mag_sus_phys_scat} (a) compares the physical quantities between the experimental and calculated results by the estimated spin Hamiltonian with $J_1 = -8.54$ meV, $J_2 = -2.67$ meV, $J_3 = -3.90$ meV, and $J_4 = 6.24$ meV.
A good fit can be obtained except for the low temperature magnetization curves.
In particular,
for 1.3 K, the fitting result is worse and the 1/4 magnetic plateau-like behavior is observed in the calculation results.
Since our target Hamiltonian only has eight spins, 
the appearance of such plateau cannot be avoided at low temperatures.
To describe the low temperature magnetizations,
the target Hamiltonian must include further magnetic interactions besides the four interactions.

Moreover, Fig.~\ref{fig:mag_sus_phys_scat} (b) shows the scatterplots of the sampling results by the same MCMC in Fig.~\ref{fig:scatter} for all combinations of errors between the experimental and calculated results, i.e., $\sum_{l=1}^L \left( y_n^{\rm ex}(g_l) - y_n^{\rm cal} (g_l, \mathbf{x}) \right)^2$.
The errors of susceptibility and magnetization curves with 50 K, 30 K, and 20 K have strong correlations.
For example,
susceptibility has a positive correlation with magnetization below 50 K,
but it has a negative correlation with magnetizations under 30 K and 20 K.
This means that if the susceptibility is fitted, the magnetization under 50 K is naturally fitted,
while the fittings of magnetizations with 30 K and 20 K become worse.
On the other hand, 
magnetization curves with 4.2 K and 1.3 K are almost independent of the other errors around the estimated magnetic interactions.
In this way,
by drawing the distributions of the magnetic interactions or physical quantities by MCMC,
their relations can be understood and the characteristics of the estimated spin Hamiltonian are extracted.

\subsection{Prediction of the magnetic properties}

\begin{figure}[]
\begin{center}
\includegraphics[scale=0.3]{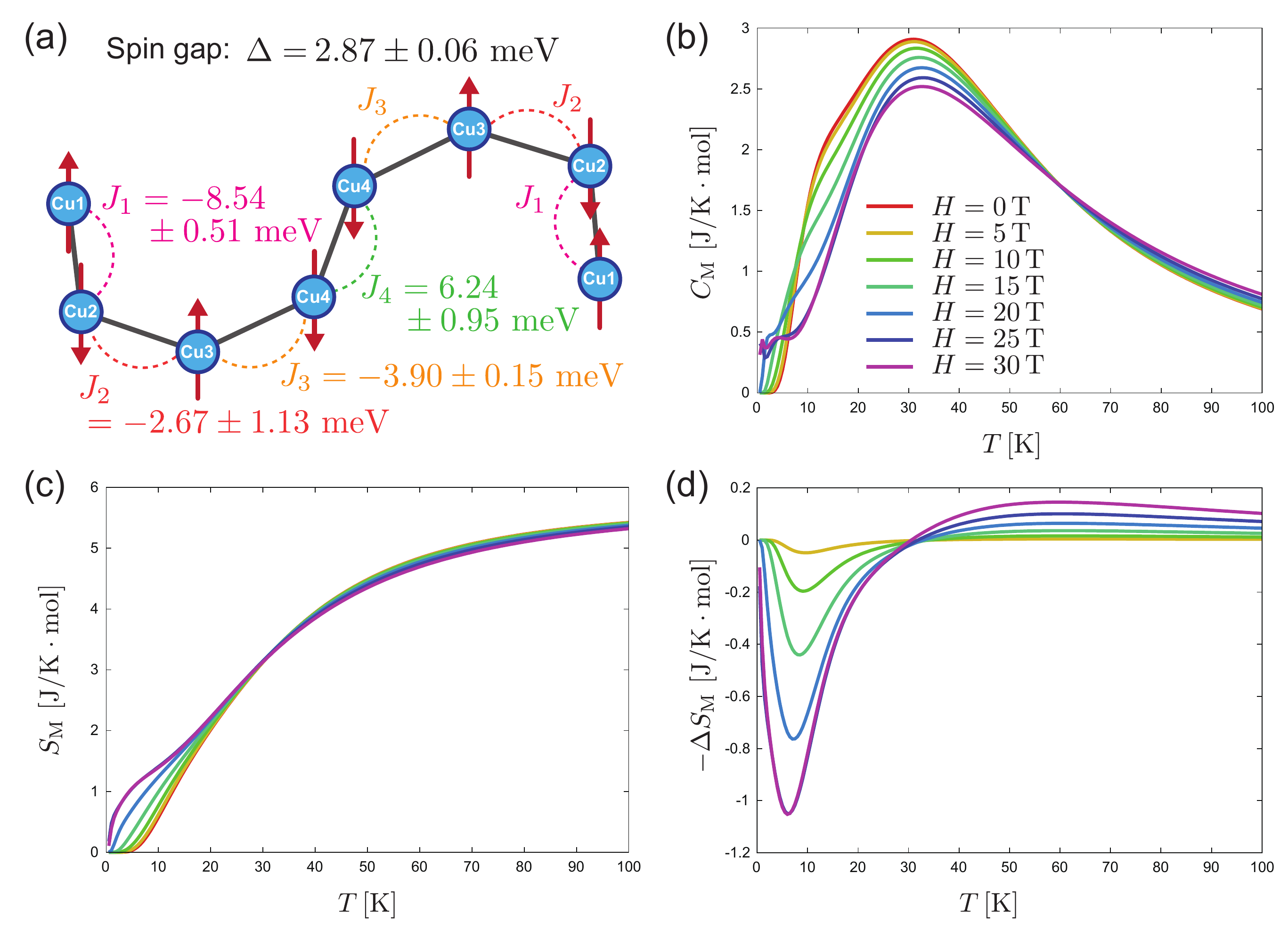} 
\end{center}
\caption{\label{fig:prediction}
(Color online)
(a) Spin gap value and spin configuration at the ground state without a magnetic field obtained by the estimated spin Hamiltonian for KCu$_4$P$_3$O$_{12}$.
Temperature dependences of (b) the magnetic specific heat, (c) magnetic entropy, and (d) magnetic entropy change when the magnetic field changes from $H$ to 0 T are plotted.
}
\end{figure}

The most important benefit from estimating the effective model is predicting magnetic properties that cannot be easily or have not been measured.
Thus, the value of the spin gap,
which is the energy gap between the ground and excited states,
the spin configuration at the ground state without a magnetic field,
and the temperature dependences of magnetic specific heat and magnetic entropy,
are calculated (Fig.~\ref{fig:prediction}).
The predictions indicate that the magnetic specific heat in KCu$_4$P$_3$O$_{12}$ will have a peak around 30 K, 
but increasing the magnetic field suppresses this peak (Fig.~\ref{fig:prediction} (b)).
In this magnet, the magnetic entropy will be almost unchanged by a magnetic field over 20 K (Fig.~\ref{fig:prediction} (c)).
Furthermore,
to predict the magnetic refrigeration property\cite{Gschneidner-2000,Tishin-2003,Gschneidner-2005,Matsumoto-2011,Franco-2012,Franco-2018}, 
the change in magnetic entropy is also evaluated (Fig.~\ref{fig:prediction} (d)).
The inverse magnetocaloric effect will be observed in KCu$_4$P$_3$O$_{12}$, 
which is a characteristic behavior of antiferromagnets\cite{Sandeman-2006,Krenke-2007,Tamura-2014,Tamura-2014b},
and the magnetic entropy change should be quite small.
In this way,
using an effective model determined by our data-driven approach,
difficult-to-measure properties can be predicted,
improving the understanding of the magnetic properties of KCu$_4$P$_3$O$_{12}$.

%

%%%%%%%%%%%%%%%%%%%%%%%%%%%%%%%%
%Discussion and summary
%%%%%%%%%%%%%%%%%%%%%%%%%%%%%%%%
\section{Discussion and summary}

We have determined the spin Hamiltonian of KCu$_4$P$_3$O$_{12}$ using a data-driven approach.
The flow of our prescription of data-driven approach is as follows.
(i) Assume a target effective model and the posterior distribution.
This constitutes the difference between the experimental and calculated results obtained by an effective model and the appropriate prior distribution of model parameters.
(ii) Determine an appropriate hyperparameter in the prior distribution by the elbow method for the MAP estimation results and estimated model parameters.
(iii) Obtain a plausible observation noise to minimize the Bayes free-energy.
(iv) Perform MCMC samplings using an estimated noise amplitude around the estimated model parameters in (ii).
From the obtained sampling data distributions,
determine the model parameters with uncertainty.
(v) Predict various properties, which cannot be easily measured in experiments using the estimated effective model.

Our data-driven approach found that the spin Hamiltonian of KCu$_4$P$_3$O$_{12}$ is the quantum Heisenberg model on the zigzag chain with eight spins of $J_1= -8.54 \pm 0.51 \ {\rm meV}$, 
$J_2 = -2.67 \pm 1.13 \ {\rm meV}$,
$J_3 = -3.90 \pm 0.15 \ {\rm meV}$,
and $J_4 = 6.24 \pm 0.95 \ {\rm meV}$.
In this estimation,
the magnetic susceptibility and magnetization curves at various temperatures are fitted.
This model can describe the experimental results with very small error, except for the extremely low temperature magnetization curves,
demonstrating that such a data-driven approach is useful to estimate the effective model.
Our approach should be  useful in parallel with ab initio calculations.

On the other hand,
the results by the data-driven approach should strongly depend on the inputted data.
Actually,
we estimated different magnetic interactions
when the inputted data is only one type of physical quantity (See Figs. S3 -- S8 in supplemental material.).
In this case, 
the fitting of the inputted quantities is well performed,
but the experimental results, 
which are not used in the estimation are not well fitted.
This fact means that each physical quantity can be well described by multiple sets of magnetic interactions.
Consequently,
preparing multiple kinds of physical quantities as the input data will not only estimate model parameters more uniquely but will also produce a more reliable effective model.

Recently, the parameter estimation for real materials by data-driven techniques has been frequently performed. 
These data driven techniques are roughly divided into two strategies. 
(i) A regression model that explains material parameters from feature variables such as material composition and structure is constructed from a large amount of collected data for known materials\cite{Pilania-2013,Ward-2016aa,Stanev-2018aa}. 
The aim of this strategy is to predict parameters in unknown materials using the trained regression model via data-driven approach.
(ii) The material parameters are determined by minimizing the discrepancy between physical quantities of the target material and those obtained by a  computational model with the material parameters\cite{Franulovic-2009,El-Naggar-2012,Tsukada-2019aa}. 
The aim of this strategy is to understand properties of the target material through the estimated parameters in the computational model.
The former, however, requires the preparation of a large amount of data sets of material compositions and structures paired with target parameters, i.e., magnetic interactions in our problem. 
Therefore, for the purpose of an estimation of magnetic interactions, 
the former is not necessarily suitable, 
but the latter is more suitable, 
and indeed our method belongs to the latter strategy.
Furthermore, our proposed noise estimation method using MCMC to evaluate the uncertainty of estimated parameters can be applied to various methods belonging to the latter strategy, 
and we believe that it is also useful in various data-driven techniques to estimate materials parameters in the computational model.

Our data-driven approach will come into its own when easy-to-measure properties obtained in the laboratory such as SQUID are inputted.
That is,
our data-driven approach can predict difficult-to-measure properties, which are obtained by large-scale experimental equipment such as neutron scattering. 
Thereby,
our data-driven approach will reduce the cost of materials development and accelerate discoveries of novel materials.

%%%%%%%%%%%%%%%%%%%%%%%%%%%%%%%%
%Acknowlegement
%%%%%%%%%%%%%%%%%%%%%%%%%%%%%%%%
\section*{Acknowledgement}

We thank Hideaki Kitazawa and Noriki Terada for the valuable discussions and Takao Furubayashi for ESR measurements.
This article is supported by
the ``Materials Research by Information Integration'' Initiative (MI2I) project
and
JST-Mirai Program Grant No. JPMJMI18A3.
M.H. was partially supported by a grant for advanced measurement and characterization technologies accelerating the materials innovation at the National Institute for Materials Science (NIMS).
The computations were performed on Numerical Materials Simulator at NIMS and the supercomputer at Supercomputer Center, Institute for Solid State Physics, The University of Tokyo.

%%%%%%%%%%%%%%%%%%%%%%%%%%%%%%%%
%References
%%%%%%%%%%%%%%%%%%%%%%%%%%%%%%%%

\providecommand{\noopsort}[1]{}\providecommand{\singleletter}[1]{#1}%

\end{document}